\documentclass[a4]{article}
\usepackage{amssymb}

\date{}

\def\ci{\perp\!\!\!\perp}

\usepackage{amssymb}

\newtheorem{teiri}{Theorem}
\newtheorem{hodai}{Lemma}
\newtheorem{rei}{Example}

\begin{document}
%
\title{Universal Bayesian Measures\\ and Universal Histogram Sequences\thanks{This paper was partially presented at IEEE International Symposium on Information Theory, Instanbul, Turkey, July, 2013.
}}
%
%
%

\author{Joe~Suzuki
\thanks{J. Suzuki is with the Department
of Mathematics, Osaka University, Toyonaka, Osaka, 560-0043, JAPAN,
e-mail: suzuki@math.sci.osaka-u.ac.jp}
}

\maketitle

\begin{abstract}
Consider universal data compression: the length $l(x^n)$ of sequence $x^n\in A^n$
 with finite alphabet $A$ and length $n$
satisfies Kraft's inequality over $A^n$, and 
$-\frac{1}{n}\log \frac{P^n(x^n)}{Q^n(x^n)}$ almost surely converges to zero as $n$ grows
for the 
$Q^n(x^n)=2^{-l(x^n)}$ and any stationary ergodic source $P$.
In this paper, we say such a $Q$ is a universal Bayesian measure.
We generalize the notion to the sources in which 
the random variables may be either discrete, continuous, or none of them. 
The basic idea is due to Boris Ryabko who utilized model weighting over histograms that approximate $P$, assuming that
a density function of $P$ exists.
However, the range of $P$ depends on the choice of the histogram sequence.
The  universal Bayesian measure constructed in this paper overcomes the drawbacks and
 has many applications to infer relation among random variables, and extends the application area of the minimum description length principle.
 
keywords: universal coding, Radon-Nikodym, Bayesian measure, estimation, density function
\end{abstract}


%

\section{Introduction}

Suppose
we wish to know if discrete random variables $X, Y$ are independent ($X\ci Y$) given $n$ pairs of examples
$\{(x_i,y_i)\}_{i=1}^n$ emitted by $(X,Y)$. 
If the probabilities of $x^n=(x_1,\cdots,x_n)$, $y^n=(y_1,\cdots,y_n)$, and $(x^n,y^n)$ are expressed by 
$P^n_X(x^n|\theta_X)$, 
$P^n_Y(y^n|\theta_Y)$, and
$P^n_{XY}(x^n,y^n|\theta_{XY})$, respectively, using unknown parameters $\theta_X,\theta_Y, \theta_{XY}$,
one way to deal with this problem 
is to decide  $X\ci Y$ if and only if
$$p Q^n_X(x^n)Q^n_Y(y^n) \geq (1-p)Q^n_{XY}(x^n,y^n)\ ,$$
 where $p$ is the prior probability of $X\ci Y$,
and the three values are defined by
\begin{eqnarray}\label{eq00}
Q^n_X(x^n):=\int P^n(x^n|\theta_X)w_X(\theta_X)d\theta_X\ , \nonumber\\
Q^n_Y(y^n):=\int P^n(y^n|\theta_Y)w_Y(\theta_Y)d\theta_Y\ , \nonumber\\
Q^n_{XY}(x^n,y^n):=\int P^n(x^n,y^n|\theta_{XY})w_{XY}(\theta_{XY})d\theta_{XY}
\end{eqnarray}
using weights $w_X,w_Y,w_{XY}$ over the parameters $\theta_X,\theta_Y,\theta_{XY}$, respectively.

To this end, let $A$ be the finite set in which $X$ takes values. 
There are many options of $Q_X$ such that 
\begin{equation}\label{eq01}
\sum_{x^n\in A^n}Q^n_X(x^n)\leq 1\ .
\end{equation}
For example\footnote{$|A|$ denotes the cardinality of set $A$.}, $Q^n_X(x^n)=|A|^{-n}$
for $x^n\in A^n$ satisfies the condition.
However, such a $Q_X$ cannot be an alternative of $P$ for large $n$ because $Q^n_X$ does not converges to $P^n$ in any sense.
On the other hand, 
if we choose 
$w_X(\theta_X)\propto \prod_{x\in A}\theta_x^{-a[x]}$ with constants $\displaystyle (a[x]=\frac{1}{2})_{x\in A}$ (Krichevsky-Trofimov \cite{kt}), then
the quantity $-\frac{1}{n}\log Q_X^n(x^n)$ almost surely converges to its entropy $H(\theta_X)$ for any 
independent and identically distributed (i.i.d) source $P^n(x^n|\theta_X)=\prod_{x\in A}\theta_x^{-c[x]}$ with parameters $\theta=(\theta_x)_{x\in A}$ and
 frequencies $(c[x])_{x\in A}$ in $x^n\in A^n$ \cite{ryabko2}.
Furthermore, the Shannon-McMillian-Breiman theorem \cite{cover} states that $\displaystyle -\frac{1}{n}\log P^n(x^n|\theta_X)$ almost surely converges to
$H(\theta_X)$ for any stationary ergodic source $\theta_X$, so that almost surely
\begin{equation}\label{eq02}
\frac{1}{n}\log \frac{P_X^n(x^n)}{Q_X^n(x^n)}\rightarrow 0
\end{equation}
if we write $P^n(x^n|\theta_X)$ by $P_X^n(x^n)$.
In this paper, we say such a $Q_X$ satisfying (\ref{eq01})(\ref{eq02}) to be a {\it universal Bayesian measure} associated with finite set $A$.
 From the above discussion, we can say that 
a universal Bayesian measure exists for finite sources. 

However, what if $X,Y$ are arbitrary without assuming they are discrete?
Recently, for random variable $X$ such that its density function $f_X$ exists,
Boris Ryabko \cite{ryabko} proved that there exists $g_X$ such that
$$
\int_{x^n\in {\mathbb R}^n} g_X^n(x^n)\leq 1
$$
and
\begin{equation}\label{eq03}
\frac{1}{n}\log \frac{f_X^n(x^n)}{g_X^n(x^n)}\rightarrow 0\ .
\end{equation}
for any $f_X$ satisfying a condition which will be specified in the later sections.

In addition, in order to decide  whether $X\ci Y$ or not is made, we need to construct
 Bayesian measures $Q_{XY}$ and $g_{XY}$ for two variables $X,Y$ extending $Q_X$ and $g_X$ for one variable $X$.

We admire Ryabko's original work \cite{ryabko}, and admit that the basic idea was already there.
However, we need to seek further generalizations for practical development of the theory.
The purposes of this paper are 
\begin{enumerate}
\item to remove the constraint that $X$ should be either discrete or continuous to obtain a general form of universality containing
 (\ref{eq02})(\ref{eq03}) as special cases; 
\item to remove the condition that Ryabko \cite{ryabko} posed; and
\item to construct universal measures for more than one variables,
\end{enumerate}
so that we establish  that a universal Bayesian measure unconditionally exists for any stationary ergodic random variable
 which may be either discrete, continuous, or none of them.
Once we can deal with universal Bayesian measures for more than one random variables, we can infer relation among them 
from given examples.

For simplicity, in this paper, we assume that the underlying source is i.i.d.
although the discussion will hold for stationary ergodic sources.

This paper is organized as follows: Section 2 gives a basic material and background of this paper.
In Sections 3,4,5, we solve the three problems above in the form of Theorems 1,2,3, respectively.
Section 6 concludes this paper by suggesting applications to show how significant the three results are.
Throughout the paper, we denote the entire real, rational, integer, and natural numbers by ${\mathbb R}, {\mathbb Q},{\mathbb Z}$, and ${\mathbb N}$, respectively.

\section{Preliminaries}

\subsection{Ryabko's measure}

Let $X$ be a random variable for which a density function $f_X$ exists, and
$A$ the set in which $X$ takes values.
Let $\{A_j\}_{j=0}^\infty$ be such that $A_0:=\{A\}$ and that
$A_{j+1}$ is a refinement of $A_j$. 
\begin{rei}\label{rei1}\rm
If\footnote{For $b<c$, $[b,c)$ denotes the set $\{x\in {\mathbb R}|b\leq x<c\}$.} $A=[0,1)$, the sequence  $A_0=\{[0,1)\}$\\
\begin{tabular}{l}
$A_1=\{[0,1/2), [1/2,1)\}$\\
$A_2=\{[0,1/4), [1/4,1/2),[1/2,3/4),[3/4,1)\}$\\
$\ldots$\\
$A_j=\{[0,2^{-(j-1)}), [2^{-(j-1)},2\cdot 2^{-(j-1)})$,\\
$\cdots, [(2^{j-1}-1)2^{-(j-1)}, 1)\}$\\
$\ldots$\\
\end{tabular}\\
satisfies the condition. 
\end{rei}

Let $\lambda$ be the Lebesgue measure.
For example, if $a=[b,c)$, then $\lambda(a)=c-b$.
For each $j=1,2,\cdots$,
let $s_j: A\rightarrow A_j$ be such that $x\in a\in A_j \Longrightarrow s_j(x)=a$,
$$P_j(a):=\int_{x\in a}f_X(x)dx\ ,\ a\in A_j$$
and
$$f_j(x):=\frac{P_j(s_j(x))}{\lambda(s_j(x))}\ ,\ x\in A\ .$$

Then, we consult the following lemma:
\begin{hodai}[\cite{ryabko2}]\rm
For any  finite set $A$,
a universal Bayesian measure associated with $A$.
\end{hodai}
(For proof, see Appendix A.)

Since each $A_j$ is a finite set, we can construct a universal Bayesian measure $Q_j$ associated with $A_j$.

Suppose we are given $x^n=(x_1,\cdots,x_n)\in A^n$ such that $(s_j(x_1),\cdots,s_j(x_n))=(a_1,\cdots,a_n)\in A_j^n$. Then, for
each $j=1,2,\cdots$, we approximate the value of the approximated 
density function
$$f^n_j(x^n):=f_j(x_1)\cdots f_j(x_n)=\frac{P_j(a_1)\cdots P_j(a_n)}{\lambda(a_1)\dots \lambda(a_n)}$$
by 
$$g_j^n(x^n):=\frac{Q^n_j(a_1,\cdots,a_n)}{\lambda(a_1)\cdots\lambda(a_n)}\ .$$

Let $\{\omega_j\}_{j=1}^\infty$ be such that $\sum \omega_j=1$, $\omega_j>0$. Ryabko proved \cite{ryabko} that
$g^n_X(x^n)=\sum_{j=0}^\infty w_jg_j^n(x^n)$ and $f_X^n(x^n)=f_X(x_1)\cdots f_X(x_n)$ satisfy (\ref{eq03})
for any $f$ such that $D(f_X||f_j)\rightarrow 0$ as $j\rightarrow \infty$,
where $D(f||g)$ is the Kullback-Leibler divergence of $g$ from $f$:
$$\displaystyle D(f||g):=\int_{x\in A} f(x)\log \frac{f(x)}{g(x)}dx\ .$$

We should notice that the set $\{f_X|D(f_X||f_j)\rightarrow 0\ {\rm as}\ j\rightarrow \infty\}$ depends on the histogram sequence $\{A_j\}$.
In this sense, Ryabko's measure is a universal Bayesian measure w.r.rt. a specific $\{A_j\}$.
In this paper, we refer this constraint to Ryabko's condition.


\subsection{Exactly when a density function exists ?}
Let $\cal B$ the entire Borel sets of  $\mathbb R$.
Formally, $X$ is a random variable if $(X\in D)\in {\cal F}$  for any $D\in {\cal B}$ for the underlying probability space $(\Omega,{\cal F},P)$.
Given random variable $X$, a necessary and sufficient condition is available that its density function exists:
\begin{rei}\rm \label{rei2}
The following two condition are equivalent \cite{billingsley}:
\begin{enumerate}
\item For any $D\in {\cal B}$, there exists $f_X: {\mathbb R}\rightarrow {\mathbb R}_{\geq 0}$ such that 
$$\mu_X(D):=P(X\in D)=\int_{x\in D}f_X(x)dx\ .$$
\item For any $D\in {\cal B}$,
$$\lambda(D):=\int_{x\in D}dx=0 \Longrightarrow \mu_X(D)=0$$
\end{enumerate}
Then, such an $f_X$ (density function) is obtained by $f_X(x)=\frac{dF_X(x)}{dx}$, where $F_X$ is the distribution function of $X$.
\end{rei}

\begin{rei}\rm \label{rei3}
We can check the following two conditions are equivalent:
\begin{enumerate}
\item For any $D\in {\cal B}$,
there exists $f_Y: {\mathbb R}\rightarrow {\mathbb R}_{\geq 0}$ such that 
$$\mu_Y(D):=P(Y\in D)=\sum_{y\in D\cap {\mathbb Z}} f_Y(y)$$
\item For any
$D\in {\cal B}$,
$$\eta(D):=|D\cap {\mathbb Z}|=0 \Longrightarrow \mu_Y(D)=0\ .$$
\end{enumerate}
Then, such an $f_Y$ is obtained by $\displaystyle f_Y(y)=P(Y=y)=\mu_Y(\{y\})$ for $y\in {\mathbb Z}$
($f_Y(y)$ may take any value for $y\not\in {\mathbb Z}$).
\end{rei}
Notice that a density function in a generalized sense exists even if the random variable is discrete.

Let $\eta$ be a $\sigma$-finite measure, i.e. there exists $\{A_j\}$  such that $A_j\in {\cal F}$, $\cup_j A_j=\Omega$ and $\eta(A_i)<\infty$ for measure space $(\Omega, {\cal F})$.
For example, $\lambda$ and $\eta$ in Examples \ref{rei2} and \ref{rei3} are both $\sigma$-finite
because $A_j=[j,j+1)$, $A_j\in {\cal B}$, $\cup_jA_j={\mathbb R}$ and $\lambda(A_j)=\eta(A_j)=1$.

For any random variable $X$ which is either discrete or continuous or none of them, there exists a density function $f_Z$ w.r.t. $\eta$ as long as 
$\mu_Z$ is absolutely continuous w.r.t. $\eta$, where $\mu_Z(D):=P(Z\in D)$ for $D\in {\cal B}$:
\begin{hodai}[Radon-Nikodym \cite{billingsley}]\label{hodai2}\rm Let $\mu, \eta$ be $\sigma$-finite measures for measure space $(\Omega, {\cal F})$. Then,
the following two conditions are equivalent:
\begin{enumerate}
\item For any $A\in {\cal F}$, 
there exists nonnegative $\displaystyle f$ such that
$\displaystyle \mu(A)=\int_Af(t)d\eta(t)$.
\item For any $A\in {\cal F}$, $\eta(A)=0 \Longrightarrow \mu(A)=0$
\end{enumerate}
\end{hodai}
If the condition in  Lemma 1 is met, we say that $\mu$ is {\it absolutely continuous} w.r.t. $\eta$ and write $\mu\ll \eta$.

We notice that the integral in the lemma is the Lebesgue integral that takes the value
$$\sup\sum_i[\inf_{\omega\in A_i}f(\omega)]\eta(A_i)$$
for $\cal F$-measurable function $f$, i.e., $\{\omega\in \Omega|f(\omega)\in D\}\in {\cal F}$ for any $D\in {\cal B}$,
where the supreme is over $\{A_i\}$ such that $A_i\cap A_j=\phi$ for $i\not=j$ and $\cup A_i=\Omega$, and contains
the Riemann integrals and summations as in Examples \ref{rei2} and \ref{rei3} as special cases.
Such a  density function $f$ in the generalized sense is called a {\it Radon-Nikodym derivative}.

\begin{rei}\rm
Let $\lambda$ and $\eta$  be as in Examples \ref{rei2} and \ref{rei3}, respectively, and $\xi(D):=\lambda(D)+\eta(D)$ for $D\in {\cal B}$.
Then, the following two conditions are equivalent:
\begin{enumerate}
\item For any $D\in {\cal B}$,
there exists $f_Z: {\mathbb R}\rightarrow {\mathbb R}_{\geq 0}$ such that 
\begin{eqnarray*}
\mu_Z(D)&:=&P(Z\in D)\\
&=&\int_{Z\in D}f_Z(z)dz+\sum_{z\in D\cap {\mathbb Z}} f_Z(z)\\
&=&\int_{z\in D} f_Z(z)d\xi(z)
\end{eqnarray*}
\item For any $D\in {\cal B}$,
$$\xi(D)=0 \Longrightarrow \mu_Z(D)=0\ .$$
\end{enumerate}
\end{rei}
Notice that a density function in the generalized sense exists even when the random variable is not either discrete or continuous.

\section{Estimating a density function in the generalized sense}

Based on the discussion in Section 2.2, we generalize the result in Section 2.1 to
the one that does not assume the random variable to be either discrete or continuous.

Let $\eta$ be a $\sigma$-finite measure.
Let $Y$ be a random variable such that
$\mu_Y\ll \eta$ for $\mu_Y(D)=P(Y\in D)$, $D\in {\cal B}$,
and $B$ the set in which $X$ takes values.
Let $\{B_k\}_{k=0}^\infty$ be such that $B_0:=\{B\}$ and that 
$B_{k+1}$ is a refinement of $B_k$. 

\begin{rei}\rm \label{rei5}
Let
$\displaystyle \eta(\{h\}):=\frac{1}{h(h+1)}$ for $h\in B={\mathbb N}=\{1,2,\cdots\}$. 
We assume that
$\mu(\{h\})>0$ only if $h\in B$. Then, $\mu\ll \eta$ for $\mu_Y(D)=P(Y\in D)$, $D\in {\cal B}$, and from Lemma \ref{hodai2},
there exists $f_Y$ such that
$$\mu_Y(D)=\sum_{h\in D}f_Y(h)\eta(\{h\})\ .$$
In fact, 
$$f_Y(h)=\frac{\mu_Y(\{h\})}{\eta(\{h\})}=h(h+1)\mu_Y(\{h\})$$
satisfies the property. For $\{B_k\}$, the following sequence satisfies the condition:\\
\begin{tabular}{l}
$B_1:=\{\{1\}, \{2,3,\cdots\} \}$\\
$B_2:=\{\{1\}, \{2\}, \{3,4,\cdots\} \}$\\
$\ldots$\\
$B_k:=\{\{1\}, \{2\},\cdots,\{k\}, \{k+1,k+2,\cdots\} \}$\\
$\ldots$\\
\end{tabular}
\end{rei}

For each $k=1,2,\cdots$, let $t_k: B\rightarrow B_k$ be such that
$y\in b\in B_k \Longrightarrow t_k(y)=b$,
$$P_k(b):=\int_{y\in b}f_Y(y)d\eta(y)$$
for $b\in B_k$, and
$$f_k(y):=\frac{P_k(t_k(y))}{\eta(t_k(y))}$$
for $y\in B$. Since $B_k$ is a finite set, we can construct a universal Bayesian measure $Q_k$ associated with
$B_k$.

Suppose we are given $y^n=(y_1,\cdots,y_n)\in B^n$ such that 
$(t_k(y_1),\cdots,t_k(y_n))=(b_1,\cdots,b_n)\in B_k^n$ for $k=1,2,\cdots$.
Then, 
for each $k=1,2,\cdots$, we estimate
$$f^n_k(y^n):=f_k(y_1)\cdots f_k(y_n)=\frac{P_k(b_1)\cdots P_k(b_n)}{\eta(b_1)\dots \eta(b_n)}\ .$$
by 
$$g_k^n(y^n):=\frac{Q^n_k(b_1,\cdots,b_n)}{\eta(b_1)\cdots\eta(b_n)}\ .$$

Let $\{\omega_k\}_{k=1}^\infty$ be such that $\sum \omega_k=1$, $\omega_k>0$. 
We claim that 
$g_X^n(y^n)=\sum_{k=0}^\infty w_kg_k^n(y^n)$ and $f_Y^n(y^n)=f_Y(y_1)\cdots f_Y(y_n)$ satisfies (\ref{eq03})
for any $f$ such that $D(f_Y||f_k)\rightarrow 0$ as $k\rightarrow \infty$, where
where $D(f||g)$ is the Kullback-Leibler divergence of $g$ from $f$:
$$\displaystyle D(f||g):=\int_{y\in B} f(y)\log \frac{f(y)}{g(y)}d\eta(y)\ .$$

For $D^n=(D_1,\cdots,D_n)$ with $D_i\in {\cal B}$,
if we define 
$$\displaystyle \mu^n_Y(D^n):=\int_{D^n} f_Y^n(y^n)d\eta^n(y^n)\ ,$$ 
$$\displaystyle \nu^n_Y(D^n):=\int_{D^n} g_Y^n(y^n)d\eta^n(y^n)\ ,$$
and
$\eta^n(D^n)=\prod_{i=1}^n\eta(D_i)$,
then we have
$$\frac{f^n(y^n)}{g^n(y^n)}= \frac{d\mu^n}{d\eta^n}(y^n)/ \frac{d\nu^n}{d\eta^n}(y^n)= \frac{d\mu^n}{d\nu^n}(y^n)\ .$$
The Kullback-Leibler  divergence of $f_k$ from $f_Y$ becomes
\begin{eqnarray*}
D(f_Y||f_k)&:=&\int f_Y(y)\log\frac{f_Y(y)}{f_k(y)}d\eta(y)\\
&=&\int \frac{d\mu_Y}{d\eta}(y)\log\{\frac{d\mu_Y}{d\eta}(y)/\frac{d\mu_k}{d\eta}(y)\}d\eta(y)\\
&=&\int d\mu_Y(y)\log\frac{d\mu_Y}{d\mu_k}(y)=D(\mu_Y||\mu_k)
\end{eqnarray*}


We arbitrarily fix $\{B_k\}_{j=0}^\infty$ so that $B_0:=\{B\}$ and that 
$B_{k+1}$ is a refinement of $B_k$. Then, the claim is stated in the following form:
\begin{teiri}\rm 
If $\mu_Y\ll \eta$, there exists a $\nu_Y^n\ll \eta^n$ such that $\nu_Y^n(B^n)\leq 1$ and 
with probability one as $n\rightarrow \infty$
\begin{equation}\label{eq20}
\frac{1}{n}\log \frac{d\mu_Y^n}{d\nu_Y^n}(y^n)\rightarrow 0
\end{equation}
for any $\mu_Y$ such that  
$D(\mu_Y||\mu_k)\rightarrow 0$ as $k\rightarrow \infty$.
\end{teiri}
Proof: First, we notice for each $y^{n-1}=(y_1,\cdots,y_{n-1})$
\begin{eqnarray*}
&&\int_{y_n\in B}g_k^{n}(y^{n})d\eta(y_n)\\
&=&g_k^{n-1}(y^{n-1})\int_{y_n\in B}\frac{Q_k^{n}(y^{n}|y^{n-1})}{\eta(t_k(y_n))}d\eta(y_n)\\
&=&g_k^{n-1}(y^{n-1})\sum_{b\in B_k}{Q_k^{n}(b|t_k(y_1),\cdots,t_k(y_{n-1}))}\\
&\leq& g_k^{n-1}(y^{n-1})
\end{eqnarray*}
Thus, 
\begin{eqnarray*}
\int_{y_n} g_Y^{n}(y^{n})d\eta(y_n)&=&\sum_k w_k \int_{y_n} g_k^{n}(y^{n})d\eta(y_n)\\
&\leq& \sum_k w_k g_k^{n-1}(y^{n-1})=g_Y^{n-1}(y^{n-1})\ ,
\end{eqnarray*}
so that $\{Z_n\}$ with
$\displaystyle z_n:=\frac{g^n_Y(y^n)}{f^n_Y(y^n)}$ is a super-martingale:
\begin{eqnarray*}
&&E[Z|y^{n-1}]\\&=&
\frac{g^{n-1}_Y(y^{n-1})}{f^{n-1}_Y(y^{n-1})}\cdot
E[\frac{g_Y^{n}(y^n)}{g_Y^{n-1}(y^{n-1})}\cdot \frac{1}{f_Y(y_n)}]\\
&\leq& z_{n-1}\cdot
\frac{1}{g_Y^{n-1}(y^{n-1})}\int_{y_n\in B}
{g_Y^{n}(y^n)}d\eta(y_n)\leq z_{n-1}\ ,
\end{eqnarray*}
where $\{Z_n\}$ is a super-martingale if and only if $\{-Z_n\}$ is a sub-martingale. From
$z_n\geq 0$, $E[Z_n]=\int g^n(y^n)d\eta^n(y^n)\leq 1$, and Doob's martingale convergence theorem below, we see
$\lim_{n\rightarrow \infty}\frac{1}{z_n}$ exists and finite,
so that with probability one as $n\rightarrow \infty$,
$$\lim_{n\rightarrow \infty}\frac{1}{z_n}>0, \lim_{n\rightarrow \infty}\log \frac{1}{z_n}>-\infty, \lim_{n\rightarrow \infty}\frac{1}{n}\log \frac{1}{z_n}\geq 0\ .$$
\begin{hodai}[Theorem 35.5 \cite{billingsley}]\rm
Let $\{X_i\}$ be a sub-martingale. If $K:=\sup_n E[|X_n|]<\infty$, then $X_n\rightarrow X$ with probability one,
where $X$ is a random variable satisfying $E[|X|]\leq 1$.
\end{hodai}
Hence
$$\lim_{n\rightarrow \infty}\frac{1}{n}\log \frac{d\mu_Y^n}{d\nu_Y^n}(y^n)\geq 0\ .$$

On the other hand, for $k=1,2,\cdots$, $\displaystyle \frac{d\nu_Y^n}{d\eta^n}(y^n)\geq w_k\frac{d\nu_k^n}{d\eta^n}(y^n)$,
so that 
\begin{eqnarray}
&&\frac{1}{n}\log \frac{d\mu_Y^n}{d\nu_Y^n}(y^n)\nonumber\\
&\leq& -\frac{1}{n}\log w_k+\frac{1}{n}\log \frac{d\mu_k^n}{d\nu_k^n}(y^n)
+\frac{1}{n}\log \frac{d\mu_Y^n}{d\mu_k^n}(y^n) \label{eq101}
\end{eqnarray}
with probability one as $n\rightarrow \infty$. Note that for each $k=1,2,\cdots$,
since $B_k$ is a finite set, there exists a universal Bayesian measure $Q_k$ associated with $B_k$, so that
$$\frac{1}{n}\log \frac{d\mu_k^n}{d\nu_k^n}(y^n)=\frac{1}{n}\log \frac{P_k^n(y^n)}{Q_k^n(y^n)}\rightarrow 0$$
with probability one as $n\rightarrow \infty$.
On the other hand, from the law of large numbers,
\begin{eqnarray*}
&&\frac{1}{n}\log \frac{d\mu_Y^n}{d\mu_k^n}(y^n)=\frac{1}{n}\sum_{i=1}^n\log \frac{d\mu_Y}{d\mu_k}(y_i)\\
&\rightarrow &E[\log \frac{d\mu_Y}{d\mu_k}]=D(\mu_Y||\mu_k)\ .
\end{eqnarray*}
with probability one as $n\rightarrow \infty$.
However, (\ref{eq101}) should hold even for large $k$. From the assumption $D(\mu_Y||\mu_k)\rightarrow 0$ as $k\rightarrow \infty$,
we require
$$\lim_{n\rightarrow \infty}\frac{1}{n}\log \frac{d\mu_Y^n}{d\nu_Y^n}(y^n)\leq 0\ .$$
This completes the proof.


\section{Universal Histogram Sequence}

Theorem 1 assumes a specific $\{B_k\}$. Given $\{B_k\}$,
for $\mu$ such that $D(\mu_k||\eta)$ does not converge to zero as $k\rightarrow \infty$,
Eq. (\ref{eq20}) does not hold in general. So, it is pleasing if we could find
$\{B_k\}$ such that for any $\mu$, $D(\mu_k||\eta)\rightarrow 0$ as  $k\rightarrow \infty$.
Then, we can remove Ryabko's condition. 

To this end, we choose any $\mu, \sigma\in {\mathbb R}$, where $\sigma$ should be positive,
and generate the following sequence:
$$C_0=\{(-\infty,\infty)\}$$
$$C_1=\{(-\infty,\mu], (\mu,\infty)\}$$
Given 
$$C_k=\{(-\infty,c_{k,1}],(c_{k,1},c_{k,2}],\cdots,(c_{k,2^k-2},c_{k,2^k-1}], (c_{k,2^k-1},\infty)\}$$
 for $k\geq 1$,
we define 
\begin{eqnarray*}
C_{k+1}&=&\{(-\infty,c_{k+1,1}],(c_{k+1,1},c_{k+1,2}],\\
&&\cdots,(c_{k+1,2^{k+1}-2},c_{k+1,2^{k+1}-1}], (c_{k+1,2^{k+1}-1},\infty)\}
\end{eqnarray*}
by
$$c_{k+1,1}=\mu-k\sigma\ ,\ c_{k+1,2^{k+1}-1}=\mu+k\sigma$$
$$c_{k+1,2j}=c_{k,j}, \ j=1,\cdots,2^{k}-1$$
$$c_{k+1,2j+1}=\frac{c_{k,j}+c_{k,j+1}}{2}\ ,\ j=1,\cdots,2^{k}-2$$

Therefore, $C_k$ contains $2^k$ elements.
In this way, given the values of $\mu,\sigma$, we obtain the sequence $\{C_k\}_{k=0}^\infty$.

Let $B$ be the set in which random variable $Y$ takes values, and define 
$$B_k^*:=\{B\cap c|c\in C_k\}\backslash \{\phi\}\ .$$

\begin{rei}\rm \label{rei6}
Let $B$ be the entire real $\mathbb R$ ($\{B_k^*\}=\{C_k\}$).
We assume that  $\eta$ is the Lebesgue measure $\lambda$, and that
$\mu_Y\ll \lambda$ for $\mu_Y(D)=P(Y\in D)$ for $D\in {\cal B}$, which means from Lemma \ref{hodai2} that 
a density function $f_Y$ exists.
Thus, for each $y\in B={\mathbb R}$, there exist a unique sequence $\{(a_k,b_k]\}_{k=1}^\infty$ such that
$-\infty<a_k\leq y\leq b_k<\infty$, $k=1,2,\cdots$, where $a_k,b_k=\pm\infty$ are allowed,
so that the ratio
$$\frac{\mu_Y((a_k,b_k]))}{\lambda((a_k,b_k])}=\frac{F_Y(b_k)-F_Y(a_k)}{b_k-a_k}$$
converges to $f(y)$ as $k\rightarrow \infty$. Thus, $D(f||f_k)\rightarrow 0$ as $k\rightarrow \infty$ for any $f$,
where $F_Y$ is the distribution function of $Y$.
\end{rei}

\begin{rei}\rm \label{rei7}
The sequence $\{B_k\}$ in Example \ref{rei5} is obtained by $\mu=1$ and $\sigma=1$ in 
the histogram sequence $\{B_k^*\}$.
Then, for each $y\in B={\mathbb N}=\{1,2,\cdots\}$, there exists $K\in {\mathbb N}$
and a unique $\{D_k\}_{k=1}^\infty$ such that $y\in D_k\in B_k^*$, $k=1,2,\cdots$ and
$\{y\}=D_k\in B_k^*$ for  $k=K,K+1,\cdots$, so that
$$f_k(y)=\frac{\mu_Y(D_k)}{\eta(D_k)}\rightarrow f(y)=\frac{\mu_Y(\{y\})}{\eta(\{y\})}$$
for each $y\in B$ and $D(f||f_k)\rightarrow 0$ as  $k\rightarrow \infty$
for any $f$.
\end{rei}

The choice of $\mu,\sigma$ may be arbitrary, but we should take the prior knowledge 
into consideration in order to make the estimation correct even for small $n$.

\begin{teiri}\rm \label{teiri}
If $\mu \ll \eta$, there exists a $\nu\ll \eta$ such that $\nu^n(B^n)\leq 1$ and 
with probability one as $n\rightarrow \infty$,
(\ref{eq20}) holds for any $\mu$.
\end{teiri}
Proof. 
It is sufficient to show $D(f||f_k)\rightarrow 0$ as $k\rightarrow \infty$ for the histogram sequence $\{B_k^*\}$.
To this end, we consult the following lemma:
\begin{hodai}[\cite{billingsley}, Problem 32.13]\rm \label{hodai3}
Let $\mu$ be the probability measure over $\cal B$, $\eta$ a $\sigma$-finite measure such that 
$\mu\ll \eta$. Then, with probability one, 
$$\lim_{h\rightarrow 0}\frac{\mu((x-h,x+h])}{\eta((x-h,x+h])}=f(x)\ ,$$
where $f$ is the density function of $\mu$ w.r.t. $\eta$.
\end{hodai}
(For proof, see Appendix B.)

In our case, for each $y\in B$, there exists  
a unique sequence $\{(a_k,b_k]\}_{k=1}^\infty$ such that
$y\in (a_k,b_k]\in B_k^*$, $k=1,2,\cdots$
and $|b_k-a_k|\rightarrow 0$ ($k\rightarrow \infty$), and obtain 
$$\lim_{k\rightarrow \infty}\frac{\mu_Y((a_k,b_k])}{\eta((a_k,b_k])}=f(y)$$
with probability one. Hence, $D(f||f_k)\rightarrow 0$ as $k\rightarrow \infty$. This completes the proof.

Hereafter, we refer $\{B_k^*\}$ to the universal histogram sequence w.r.t. $B$.

\section{When more than one variable exist}

Analogous to the one variable case, we apply the notion of estimating Radon-Nikodym derivatives
to the two random variables case.

Let $\mu_X(D_X):=P(X\in D_X)$ and $\mu_Y(D_Y):=P(Y\in D_Y)$ for $D_X,D_Y\in {\cal B}$,
and $\eta_X,\eta_Y$ $\sigma$-finite measures such that $\mu_X\ll\eta_X$ and $\mu_Y\ll \eta_Y$, respectively.
Then, for\footnote{$A\times B$ denotes the Cartesian product of sets $A,B$.} $\mu_{XY}(D_X\times D_Y)=P(X\in D_X,Y\in D_Y)$,
$D_X,D_Y\in {\cal B}$, we have $\mu_{XY}\ll \eta_X\times \eta_Y$, where
$\eta_X\times \eta_Y$ is the product measure of $\eta_X,\eta_Y$: $\eta_{X}\times\eta_Y(D_X\times D_Y)=\eta_X(D_X)\eta_Y(D_Y)$.
Hence, from Lemma \ref{hodai2}, there exists $f_{XY}$ such that
$$\mu_{XY}(D_X\times D_Y)=\int_{x\in D_x}\int_{y\in D_Y}f_{XY}(x,y)d\eta_X(x)d\eta_Y(y)\ .$$

Let $\{A_j\}, \{B_k\}$ be such that $A_0=\{A\}$ and $B_0=\{B\}$ and $A_{j+1},B_{k+1}$ are refinements of $A_j,B_k$,
where $A,B$ are the sets in which $X,Y$ take values, respectively.

For each $j,k=1,2,\cdots$, $s_j:A\rightarrow$ and $t_k:B\rightarrow B_k$ be such that
$x\in a\in A_j\Longrightarrow s_j(x)=a$ and
$y\in a\in B_k\Longrightarrow t_k(y)=b$, respectively,
\begin{eqnarray*}
P_{jk}(a,b)&:=&\int_{x\in a}\int_{y\in b}f_{XY}(x,y)d\eta_X(x)d\eta_Y(y)\\
\end{eqnarray*}
for $a\in A_j$ and $b\in B_k$, and
$$f_{jk}(x,y):=\frac{P_{jk}(s_j(x),t_k(y))}{\eta_X(s_j(x)\eta_Y(t_k(y)))}$$
for $x\in A$ and $y\in B$.

Since $A_j\times B_k$ is a finite set, we can construct a universal Bayesian measure $Q_{j,k}$ associated with 
$A_j\times B_k$.

Suppose we are given 
$x^n=(x_1,\cdots,x_n)\in A^n$ and 
$y^n=(x_1,\cdots,x_n)\in A^n$ such that 
$(s_j(x_1),\cdots,s_j(x_n))=(a_1,\cdots,a_n)$ and 
$(t_k(y_1),\cdots,t_k(y_n))=(b_1,\cdots,b_n)$
for $j,k=1,2,\cdots$.
Then, for each $j,k=1,2,\cdots$, we estimate
\begin{eqnarray*}
f_{j,k}^n(x^n,y^n)&=&f_{i,j}(x_1,y_1)\cdots f_{j,k}(x_n,y_n)\\
&=&
\frac{P_{j,k}(a_1,b_1)\cdots P_{j,k}(a_n,b_n)}{\eta_X(a_1)\cdots \eta_X(a_n)\eta_Y(b_1)\cdots \eta_Y(b_n)}
\end{eqnarray*}
by
$$g_{j,k}^n(x^n,y^n)=\frac{Q^n_{j,k}(a_1,b_1,\cdots,a_n,b_n)}{\eta_X(a_1)\cdots \eta_X(a_n)\eta_Y(b_1)\cdots \eta_Y(b_n)}\ .$$
Then, we obtain
$$f_Y^n(x^n,y^n)=f_Y(x_1,y_1)\cdots f_Y(x_n,y_n)$$
and 
$$g^n_Y(x^n,y^n)=\sum_{j,k}w_{j,k}g_{j,k}(x^n,y^n)$$
for some $\{w_{j,k}\}$  such that $w_{j,k}>0$ and $\sum_{j,k}w_{j,k}=1$.

As $\{A_j\times B_k\}$, we use $\{(A_j\times B_k)^*\}$ defined by
$$(A_j\times B_k)^*:=\{(A\cap c)\times (B\cap d)|c\in C_j, d\in C_k\}\backslash \{\phi\}$$


\begin{teiri}\rm
Suppose $\mu_X\ll \eta_X$ and $\mu_Y\ll \eta_Y$, there exists $\nu^n_{XY}\ll \eta_X^n\times \eta_Y^n$
such that $\nu^n(A^n\times B^n)\leq 1$ and with probability one as $n\rightarrow \infty$
$$\frac{1}{n}\log \frac{d\mu_{XY}^n}{d\nu_{XY}^n}(x^n,y^n)\rightarrow 0$$
for any $\mu_{XY}$.
\end{teiri}
Proof: First, we notice for each $x^{n-1}=(x_1,\cdots,x_{n-1})$ and $y^{n-1}=(y_1,\cdots,y_{n-1})$
\begin{eqnarray*}
\int_{x_n\in A}\int_{y_n\in B}
g_{j,k}^{n}(x^n,y^{n})d\eta_X(x_n)d\eta_Y(y_n)\leq g_{j,k}^{n-1}(x^{n-1},y^{n-1})
\end{eqnarray*}
By weighting by $\sum_{j,k} w_{j,k}[\cdot]$ for the both sides, we have
\begin{eqnarray*}
\int_{y_n} g_{XY}^{n}(x^n,y^{n})d\eta_X(x_n)d\eta_Y(y_n)\leq g_{XY}^{n-1}(x^{n-1},y^{n-1})\ ,
\end{eqnarray*}
so that $\{Z_n\}$ with
$\displaystyle z_n:=\frac{g^n_{XY}(x^n,y^n)}{f^n_{XY}(x^n,y^n)}$ is a super-martingale:
\begin{eqnarray*}
&&E[Z|x^{n-1},y^{n-1}]\\
&=&
\frac{g^{n-1}_{XY}(x^{n-1},y^{n-1})}{f^{n-1}_{XY}(x^{n-1},y^{n-1})}\cdot
E[\frac{g_{XY}^{n}(x^n,y^n)}{g_{XY}^{n-1}(x^{n-1},y^{n-1})}\cdot \frac{1}{f_{XY}(x_n,y_n)}]\\
&\leq &z_{n-1}\cdot
\frac{1}{g_{XY}^{n-1}(x^{n-1},y^{n-1})}\\
&&\cdot \int_{x_n\in A}\int_{y_n\in B}
{g_{XY}^{n}(x^n,y^n)}d\eta_X(x_n)d\eta_Y(y_n)\\&\leq& z_{n-1}\ .
\end{eqnarray*}
From $z_n\geq 0$, $E[Z_n]=\int g_{XY}^n(x^n,y^n)d\eta_X(x^n)d\eta_Y^n(y^n)\leq 1$, and Doob's martingale convergence theorem, we see
$\lim_{n\rightarrow \infty}\frac{1}{z_n}$ exists and finite,
so that with probability one as $n\rightarrow \infty$,
$$\lim_{n\rightarrow \infty}\frac{1}{z_n}>0, \lim_{n\rightarrow \infty}\log \frac{1}{z_n}>-\infty, \lim_{n\rightarrow \infty}\frac{1}{n}\log \frac{1}{z_n}\geq 0\ .$$
Hence
$$\lim_{n\rightarrow \infty}\frac{1}{n}\log \frac{d\mu_{XY}^n}{d\nu_{XY}^n}(x^n,y^n)\geq 0\ .$$

On the other hand, for $j,k=1,2,\cdots$, 
$$\displaystyle \frac{d\nu_{XY}^n}{d\eta_X^nd\eta_Y^n}(x^n,y^n)\geq w_{j,k}\frac{d\nu_{j,k}^n}{d\eta_X^nd\eta_Y^n}(x^n,y^n)\ ,$$
so that 
\begin{eqnarray*}
&&\frac{1}{n}\log \frac{d\mu_{XY}^n}{d\nu_{XY}^n}(x^n,y^n)\\
&\leq& -\frac{1}{n}\log w_{j,k}+\frac{1}{n}\log \frac{d\mu_{j,k}^n}{d\nu_{j,k}^n}(x^n,y^n)\\
&&+\frac{1}{n}\log \frac{d\mu_{XY}^n}{d\mu_{j,k}^n}(x^n,y^n)
\end{eqnarray*}
with probability one as $n\rightarrow \infty$. Note that for each $j,k=1,2,\cdots$,
since $A_j\times B_k$ is a finite set, there exists a universal Bayesian measure $Q_{j,k}$ associated with $A_j\times B_k$, so that
$$\frac{1}{n}\log \frac{d\mu_{j,k}^n}{d\nu_{j,k}^n}(x^n,y^n)=\frac{1}{n}\log \frac{P_{j,k}^n(x^n,y^n)}{Q_{j,k}^n(x^n,y^n)}\rightarrow 0$$
with probability one as $n\rightarrow \infty$.
On the other hand,
\begin{eqnarray*}
&&\frac{1}{n}\log \frac{d\mu_{XY}^n}{d\mu_k^n}(x^n,y^n)=
\frac{1}{n}\sum_{i=1}^n\log \frac{d\mu_{XY}}{d\mu_{j,k}}(x_i,y_i)\\
&\rightarrow &E[\log \frac{d\mu_{XY}}{d\mu_{j,k}}]=D(\mu_{XY}||\mu_{j,k})\ .
\end{eqnarray*}
So, it is sufficient to show $D(\mu_{XY}||\mu_{j,k})\rightarrow 0$ as $j,k\rightarrow \infty$ for 
histogram sequence $\{(A_j\times B_k)^*\}$.

To this end, we consult the following lemma:
\begin{hodai}\rm \label{hodai5}
Let $\mu_{XY}$ be the probability measure over ${\cal B}^2$, $\eta_X,\eta_Y$ $\sigma$-finite measures such that 
$\mu_{XY}\ll \eta_X$ and $\mu_{XY}\ll \eta_Y$. Then, with probability one, 
\begin{eqnarray*}
&&\lim_{h_x\rightarrow 0}\lim_{h_y\rightarrow 0}\frac{\mu_{XY}([x-h_x,x+h_x]\times [y-h_y,y+h_y])}{\eta_X((x-h_x,x+h_x])\eta_Y((y-h_y,y+h_y])}\\
&=&f_{XY}(x,y)\ ,
\end{eqnarray*}
where $f_{XY}$ is the density function of $X,Y$.
\end{hodai}
(For proof, see Appendix B.)

In our case, for each $(x,y)\in A\times B$, there exist a unique pair of sequences 
$\{(a_j,b_j]\}_{j=1}^\infty$ and $\{(c_k,d_k]\}_{k=1}^\infty$ such that
$x\in (a_j,b_j]\in A_j^*$, $j=1,2,\cdots$,
$y\in (c_k,d_k]\in B_k^*$, $k=1,2,\cdots$,
$|b_j-a_j|\rightarrow 0$ ($j\rightarrow \infty$),
$|d_k-c_k|\rightarrow 0$ ($k\rightarrow \infty$)
 and obtain 
$$\lim_{j\rightarrow \infty}\lim_{k\rightarrow \infty}\frac{\mu_{XY}((a_j,b_j]\times (c_k,d_k])}{\eta_X((a_j,b_j])\eta_Y((c_k,d_k])}=f_{XY}(x,y)$$
with probability one. Hence, $D(f||f_{j,k})\rightarrow 0$ as $j,k\rightarrow \infty$. Thus,
$$\lim_{n\rightarrow \infty}\frac{1}{n}\log \frac{d\mu_{XY}^n}{d\nu_{XY}^n}(x^n,y^n)\leq 0\ .$$
This completes the proof.

\begin{rei}\rm
Suppose $\{A_j\}, \{B_k\},\eta_X,\eta_Y$ are given by the universal histogram sequences w.r.t. $A=[0,1),B=\{1,2,\cdots\}$ and  $\eta_X=\lambda, \eta_Y=\eta$
in Examples \ref{rei1} and \ref{rei5}. Then, we can construct $\frac{d\nu_{XY}^n}{d\eta_X^nd\eta_Y}(x^n,y^n)$ from $x^n\in A^n$ and $y^n\in B^n$.
\end{rei}

It is straightforward to extend the result for the two variable case to the $m (\geq 2)$ variables case.

\section{Concluding Remarks}

In this paper, we successfully construct a universal Bayesian measure for any random variables:
\begin{enumerate}
\item we extended Ryabko's measure so that the random variables may be either discrete or continuous or none of them; 
\item constructed a universal histogram sequence that realizes universality for any source; and
\item constructed a universal Bayesian measure for more than one variables.
\end{enumerate}

The results in this paper are rather theoretical but contain many applications such as
\begin{enumerate}
\item Bayesian network structure learning \cite{uai,dcc},
\item a variant of the Chow-Liu algorithm learning a forest given examples \cite{uai,pgm}\ .
\end{enumerate}
In fact, in any database, both discrete and continuous fields are present. Then, we need to find dependency 
among those attributes. However, the existing results only dealt with either only discrete data or only continuous data.
This paper deals with the most general and realistic cases. 

For contributions to statistics, constructing such a universal Bayesian measure means
  establishing  a general form of Bayesian Information Criteria (BIC).
Suppose we have a countable number of models $m=1,2,\cdots$ each of which expresses a relation among 
random variables. If we construct a universal Bayesian measure $q(x^n|m)$ w.r.t. model $m$ given data $x^n$, then 
we can select $m$ such that  $-\log p(m)-\log q(x^n|m)$ is minimized, where $p(m)$ is the prior probability of model $m$. In fact,
the measure applies to all the cases that BIC/MDL applied thus far.

\section*{Appendix A: Proof of existence of $Q$ satisfying (2) and (3) for finite set $A$}
Although the proposition is standard \cite{ryabko2}, we give a proof for selfcontainedness.

Let $c[x]$ be the frequency of $x\in A$ in $x^n=(x_1,\cdots,x_n)\in A^n$.
Then, we see 
$$Q^n_X(x^n):= 
\frac
{\Gamma(\sum_{x\in A}a[x])\prod_{x\in A}(c[x]+a[x])}
{\Gamma(\sum_{x'\in A} (c[x']+a[x'])) \prod_{x\in A} \Gamma(a[x])}$$
with $a[x]=\frac{1}{2}$, $x\in A$,  satisfies (2), where $m=|A|$ and $\Gamma$ is the Gamma function: $\Gamma(z)=\int_0^\infty t^{z-1}e^{-t}dt$.
In fact, $Q_X^0(x^0)=1$; and if (2) is assumed for $n\geq 0$, then for $x_{n+1}\in A$ and $x^{n+1}=(x_1,\cdots,x_{n+1})\in A^{n+1}$,
\begin{eqnarray*}
&&\sum_{x^{n+1}\in A^{n+1}}Q^{n+1}(x^{n+1})\\
&=&\sum_{x^{n+1}\in A^{n+1}}Q_X^n(x^n)\cdot \frac{c[x_{n+1}]+\frac{1}{2}}{n+\frac{m}{2}}\\
&=&\sum_{x^n\in A^n}Q_X^n(x^n)\sum_{x_{n+1}\in A}\frac{c[x_{n+1}]+\frac{1}{2}}{n+\frac{m}{2}}\leq 1\ ,
\end{eqnarray*}
where $\Gamma(z+1)=z\Gamma(z)$ has been applied for $z>0$. Thus, we obtain (2).

From Stirling's formula\footnote{The natural logarithm is assumed.}:
$$\log \Gamma(z)=(z-\frac{1}{2})\log z-z-\frac{1}{2}\log (2\pi)+o(1)\ ,$$
we have 
\begin{eqnarray*}
&&-\log Q^n_X(x^n)\\
&=&(n+\frac{m-1}{2})\log(n+\frac{m}{2})-(n+\frac{m}{2})\\
&&-\sum_{x\in A}c[x]\log (c[x]+\frac{1}{2})+\sum_{x\in A}(c[x]+\frac{1}{2})\\
&&+\frac{m-1}{2}\log (2\pi)+o(1)\\
&=&-\sum_{x\in A}c[x]\log \frac{c[x]+\frac{1}{2}}{n+\frac{m}{2}}+\frac{m-1}{2}\log (x+\frac{m}{2})+O(1)
\end{eqnarray*}
From the law of large numbers,  $c[x]/n$ converges to the probability $P(x)$ of $x\in A$ with probability one as $n\rightarrow \infty$ for independent
source $P_X^n(x^n)=\prod_{i=1}^nP(x_i)$ for $x^n=(x_1,\cdots,x_n)\in A^n$, so
that $-\log Q_X^n(x^n)$ converges to its entropy $H(P):=\sum_{x\in A}-P(x)\log P(x)$.
On the other hand, from the law of large numbers, with probability one as $n\rightarrow \infty$
\begin{eqnarray*}
&&-\frac{1}{n}\log P_X^n(x^n)
=\frac{1}{n}\sum_{i=1}^n\{-\log P(x_i)\}\\
&\rightarrow &E[-\log P(X)]=H(P)
\end{eqnarray*}
(Shannon-McMillian-Breiman \cite{cover}). Thus, we obtain (3), and this completes the proof.

\section*{Appendix B: Proof of Lemmas 4 and 5}
For the one variable case, let $y\in {\mathbb R}$.
Since $\mu_Y\ll \eta$ for $\mu_Y(D)=P(Y\in D)$, $D\in {\cal B}$, 
there exists $f_Y: {\mathbb R}\rightarrow {\mathbb R}_{\geq 0}$ such that for $h>0$,
$$\mu_Y([y-h,y+h])=\int_{y-h}^{y+h}f_Y(u)d\eta(u)$$
Let $\varphi: {\mathbb R}\rightarrow {\mathbb R}$ be such that
$$\varphi_Y(t):=inf\{z\in {\mathbb R}|t\leq \eta^*([y,z])\}\ ,$$ where 
$$\eta^*([y,z]):=\left\{
\begin{array}{ll}
\eta([y,z]),& y\leq z\\
-\eta([z,y]),& y>z
\end{array}\ .
\right.$$
Then, the integral is expressed by 
$$\int_{-\eta[y-h,y]}^{\eta[y,y+h]}f_Y(\varphi_Y(t))dt$$
On the other hand, in general, for density function $f$ and  $x\in {\mathbb R}$, we have
$$\lim_{h\rightarrow 0} \frac{\int_{x-h}^{x+h}f(t)dt}{2h}=f(x)$$
Thus, as $h\rightarrow 0$, we have
\begin{eqnarray*}
&&\frac{\mu_Y([y-h,y+h])}{\eta([y-h,y+h])}
=\frac{\int_{-\eta[y-h,y]}^{\eta[y,y+h]}f_Y(\varphi_Y(t))dt}{\eta([y,y+h])-\{-\eta([y-h,y])\}}\\
&\rightarrow& f_Y\circ\varphi(0)=f_Y(y)
\end{eqnarray*}

For the two variable case, let $x,y\in {\mathbb R}$.
Since $\mu_{XY}\ll \eta_X,\eta_Y$ for $\mu_{XY}(D)=P(X\in D_X,Y\in D_Y)$, $D_X,D_Y\in {\cal B}$,
there exists $f_{XY}: {\mathbb R}\rightarrow {\mathbb R}_{\geq 0}$ such that for $h_x,h_y>0$
\begin{eqnarray*}
&&\mu_{XY}([x-h_x,x+h_x]\times [y-h_y,y+h_y])\\
&=&\int_{x-h_x}^{x+h_x}\int_{y-h_y}^{y+h_y}f_{XY}(u,v)d\eta_X(u)d\eta_Y(v)
\end{eqnarray*}
Let $\varphi: {\mathbb R}^2\rightarrow {\mathbb R}^2$ be such that
\begin{eqnarray*}
&&\varphi_Y(s,t)\\
&:=&(inf\{w\in {\mathbb R}|s\leq \eta^*([x,w])\},inf\{z\in {\mathbb R}|t\leq \eta^*([y,z])\})\ .
\end{eqnarray*}
Then, the integral is expressed by 
$$\int_{-\eta_X[x-h_x,x]}^{\eta_X[x,x+h_x]}\int_{-\eta_Y[y-h_y,y]}^{\eta_Y[y,y+h_y]}f_{XY}(\varphi_{XY}(s,t))dsdt$$
On the other hand, in general, for density function $f$ and  $x,y\in {\mathbb R}$, we have
$$\lim_{h_x\rightarrow 0}\lim_{h_y\rightarrow 0} \frac{\int_{x-h_x}^{x+h_x}\int_{y-h_y}^{y+h_y}f(s,t)dsdt}{4h_xh_y}=f(x,y)$$
Thus, as $h_x,h_y\rightarrow 0$, we have
\begin{eqnarray*}
&&\frac{\mu_Y([x-h_x,x+h_x]\times [y-h_y,y+h_y])}{\eta_X([x-h_x,x+h_x])\eta_Y([y-h_y,y+h_y])}\\
&=&\{\int_{-\eta_X[x-h_x,x]}^{\eta_X[x,x+h_x]}\int_{-\eta_Y[y-h_y,y]}^{\eta_Y[y,y+h_y]}f_{XY}(\varphi_{XY}(s,t))dsdt\}\\
&&/ \{[\eta_X([x,x+h_x])-\{-\eta_X([x-h_x,x])\}]\\
&&\cdot [\eta_Y([y,y+h_y])-\{-\eta_Y([y-h_y,y])\}]\}\\
&\rightarrow& f_{XY}\circ\varphi(0,0)=f_{XY}(x,y)
\end{eqnarray*}
This completes the proof.

\section*{Acknowledgment}The authors would like to thank Prof. Boris Ryabko for suggesting me to
 develop further applications of his exciting theory. 

\end{document}